\documentclass[aps,prb,superscriptaddress,preprint,showpacs]{revtex4}

\usepackage{graphicx}
\usepackage{rotating}
\usepackage{longtable} 
\usepackage{SIunits}
\usepackage[dvips]{color}

\begin{document} 

\newcommand{\lsmo}{La$_{0.5}$Sr$_{1.5}$MnO$_4$}
\newcommand{\oo}{($\frac{1}{4}$,$\frac{1}{4}$,0)}
\newcommand{\mo}{($\frac{1}{4}$,$-\frac{1}{4}$,$\frac{1}{2}$)}

\title{Resonant Soft X-ray Scattering Investigation of Orbital and Magnetic Ordering in $\rm La_{0.5}Sr_{1.5}MnO_4$}
\author{S. B. Wilkins} 
\affiliation{European
  Commission, Joint Research Center, Institute for Transuranium Elements,
Hermann von Helmholtz-Platz 1, 76344 Eggenstein-Leopoldshafen,
  Germany}
\affiliation{European Synchrotron Radiation
  Facility, Bo\^\i te Postal 220, F-38043 Grenoble Cedex, France}
\author{N. Stoji\' c}
\affiliation{Abdus Salam International Centre for Theoretical Physics, Trieste 34014, Italy}
\author{ T. A. W. Beale}
\affiliation{Department of Physics, University of Durham,
Rochester Building, South Road, Durham, DH1 3LE,
UK}
\author{ N. Binggeli}
\affiliation{Abdus Salam International Centre for Theoretical Physics, Trieste 34014, Italy}
\author{C. W. M. Castleton}
\affiliation{Materials and Semiconductor Physics Laboratory,
Royal Institute of Technology (KTH), Electrum 229, 16440 Kista,
Sweden}
\affiliation{Department of Physical Electronics/Photonics, ITM, Mid Sweden University, 85170 Sundsvall, Sweden}
\author{P. Bencok}
\affiliation{European Synchrotron Radiation
  Facility, Bo\^\i te Postal 220, F-38043 Grenoble Cedex, France}
\author{D. Prabhakaran}
\author{A. T. Boothroyd}
\affiliation{Department of Physics, University of Oxford,
Clarendon Laboratory, Parks Road, Oxford, OX1 3PU, UK}
\author{P. D. Hatton}
\affiliation{Department of Physics, University of Durham,
Rochester Building, South Road, Durham, DH1 3LE,
UK}
\author{M. Altarelli}
\affiliation{Abdus Salam International Centre for Theoretical Physics, Trieste 34014, Italy}
\affiliation{Sincrotrone Trieste, Area Science Park, 34012 Basovizza, Trieste, Italy}

\date{\today}

\begin{abstract}
We report resonant x-ray scattering data of the orbital and magnetic ordering at low temperatures at the Mn $L_{\rm 2,3}$ edges in \lsmo. The orderings display complex energy features close to the Mn absorption edges. Systematic modeling with 
atomic multiplet crystal field calculations was used to extract meaningful information 
regarding the interplay of spin, orbital and Jahn-Teller order. These calculations provide 
a good general agreement with the observed energy dependence of the scattered intensity { for a} 
dominant orbital ordering of the $ d_{x^2 - z^2} / d_{y^2 - z^2} $ type. In addition, the 
origins of various spectral features are identified.   The temperature dependence of the orbital and magnetic ordering was measured and displays a strong interplay between the magnetic and orbital order parameters.
\end{abstract}


\maketitle

\section{Introduction}
 \begin{figure}
  \begin{center}
  \includegraphics[width=7.7cm]{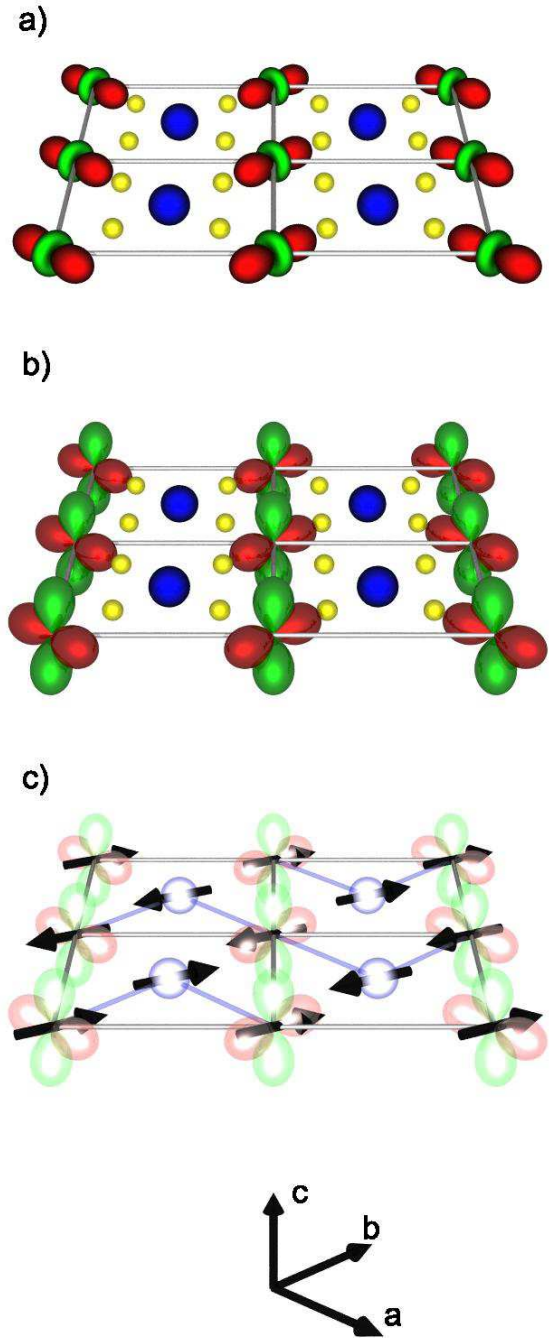}
  \caption{ The structure of \lsmo. a) and b) show two different types of orbital ordering,  
${d_{3x^2-r^2} / d_{3y^2 - r^2}}$ type
in a) and ${d_{x^2 - z^2} /d_{ y^2 - z^2}}$ in b). c) shows the magnetic ordering 
with blue lines representing the
ferromagnetic spin chains. The orbitals are drawn for the Mn$^{3+}$ ions,
while the small yellow circles represent oxygen atoms and large blue circles represent the Mn$^{4+}$ ions.  }
  \bigskip
  \label{fig:structure}
  \end{center}
\end{figure} 

Orbital ordering, which involves correlations between the spatial
distribution of the outermost valence electrons, has long been
considered a vital ingredient in the structural and physical
properties of strongly correlated electron systems such as
transition metal oxides. The competition and cooperation between
the charge, orbital, and spin degrees of freedom of the electrons
manifest itself in unusual properties such as high temperature
superconductivity, colossal magnetoresistance and
magnetostructural transitions \cite{science}. In particular,
charge-orbital ordering in half-doped manganites has attracted
much attention and controversy
\cite{radaelli:3015,mutou:3685,khomskii:3329,hotta:2478,daoud-aladine:097205}.
In \lsmo\ the Mn sites are all
crystallographically equivalent at room temperature with an average valency of +3.5.
This material displays a phase transition at $\sim$~240 K below which
it was believed that charge
disproportionation of the Mn ions occurs \cite{moritomo:3297,sternlieb:2169}, 
creating two inequivalent sites identified as Mn$^{3+}$ and Mn$^{4+}$. 
This was originally predicted by Goodenough \cite{goodenough:564}.
It should be noted that the picture of integral charge on the Mn ions
has recently been challenged \cite{HerGarSub04}, implying that the actual
valence difference of the two Mn ions is much less then 1. The assumption
that they are close to 3.5+, however, is based on the formal valence and 
does not take covalency into account  which should cause a reduction; according
to Hartree-Fock calculations on the related compound ${\rm La_{0.5}Ca_{0.5}MnO_3}$ \cite{FerTowLit03}
the charge on the manganese ion is even below 3+. In
this paper, we adopt a multiplet picture which applies to integer
charge only and we assume the value of 3+ as the closest to the real situation.
The strong Hund's
rule coupling and the  cubic ($O_h$) component of the crystal
field implies that the Mn$^{3+}$ $3d^4$ site has three electrons
in the $t_{2g\uparrow}$ level and one electron in the twofold
degenerate $e_{g\uparrow}$ level. The degeneracy of the $e_{g\uparrow}$ level
can be lifted by cooperative Jahn-Teller distortions of the
$\mathrm{MnO_6}$ octahedra reducing the symmetry to $D_{4h}$.
 Adding the tetragonal field, $t_{2g\uparrow}$ level splits into $d_{xy}$ and doubly-degenerate $d_{xz,yz}$
orbitals, while
 $e_{g\uparrow}$ level splits into ${d_{3z^2-r^2}}$ and ${d_{x^2-y^2}}$.
Therefore, depending on the value of the applied tetragonal crystal field, 
 Mn$^{3+}$  will have as a ground state
filled $t_{2g\uparrow}$ levels and the $e_{g\uparrow}$ orbital lying along the elongation direction,
either  ${d_{3z^2-r^2}}$ or ${d_{x^2-y^2}}$
orbital (see Fig~\ref{fig:structure}a and \ref{fig:structure}b respectively), which will determine the type of the orbital ordering. Further cooling below T$_{N}\sim120$~K
results in long range anti-ferromagnetic ordering of 
manganese ions into a CE type structure~\cite{wohlan:545}.
This was the basis of an alternative model put forward
by Goodenough\cite{goodenough:564} in which orbital order
correlations lower the energy by favoring  anti-ferromagnetic interactions.
Until very recently, detailed investigations of the origin of 
the orbital order remained elusive because of the lack of a technique capable of
direct observation.

Resonant x-ray scattering (RXS) studies of \lsmo\ were first attempted at the manganese
$K$ edge. The diffraction intensity at an orbital ordering reflection
displayed a striking resonant enhancement near the absorption edge and a specific 
dependence on the azimuthal angle\cite{murakami:1932}. However, the Mn~$K$ edge
resonance involves virtual excitations from $1s$ to $4p$ bands
and is thus generally insensitive to orbital ordering of the $3d$ states. 
Theoretical studies proposed that the observed
sensitivity is largely due to Jahn-Teller distortions and 4$p$
band structure effects \cite{mizokawa:r493,elfimov:4264,benfatto:636,solovyev:2825,mahadevan:066404,benedetti:060408}, rather than $3d-4p$ Coulomb interactions~\cite{IshMae98}
and this conclusion was recently supported by experimental work on strained 
manganite films \cite{OhsMurKiy03}.
Experiments performed at the $L$ edges, on the other hand, directly probe the $3d$ states.
Resonant soft x-ray diffraction at the
Mn~$L$ edges was first reported by Wilkins \emph{et
al.}~\cite{wilkins:187201} and was followed by the first direct observations
of orbital ordering using soft x-rays in
\lsmo~\cite{wilkins:167205}. Aided by earlier theoretical
predictions~\cite{castleton:1033} they reported energy resonances at the Mn~$L_{\rm 3}$ and $L_{\rm 2}$ edges
showing that the orbital ordering was caused by a mixture of both
cooperative Jahn-Teller distortions and direct Goodenough orbital
correlations. Similar experimental results were
later independently reported by another group~\cite{dhesi:056403}.
More recently, the same technique has been used to study magnetic and orbital correlations in 
${\rm Pr_{0.6}Ca_{0.4}MnO_3}$\cite{ThoHilGre04}.

In this paper, we report both experimental and theoretical 
results on resonant x-ray scattering from orbital and magnetic order
at low temperatures in \lsmo.  
We describe the measured intensity of the 
specific peaks in the energy spectra for both the magnetic and orbital reflections. We also report 
calculations using an atomic multiplet picture
incorporating the crystal field effects. We present 
the  spectra obtained by fitting the crystal 
field parameters and discuss their agreement with the
experimental spectra. Further, we indicate 
the dominant orbital ordering type.
We also present analysis of the superlattice reflections
in terms of the Jahn-Teller {\it vs.} orbital ordering origin
of the spectral features. 
Finally, we have measured the temperature dependence of certain features in 
the energy resonant spectrum. 
These show a dramatic change at a temperature corresponding to the 
N\'eel temperature, indicative of a strong interplay between 
the magnetic and orbital order parameters. 

This paper is organized as follows: Section II of the paper describes 
the methods to obtain the spectra, experimental and theoretical. The results
are presented in section III, where experimental spectra and theoretical fits are compared. There we also
present the analysis of the Jahn-Teller contributions to the orbital ordering scattering. Section IV describes
the temperature dependence of the spectra, while all our results are summarized in the conclusions.

\section{ Method}

\subsection{ Experimental approach}

The experiments were performed at beamlines 5U1 at Daresbury and
ID08 at the European Synchrotron Radiation Facility (ESRF).
Single crystals of \lsmo\ with dimensions $10\times3\times3$~mm$^{3}$
were grown at the University of Oxford using the floating zone
method. They were cut with either [110] or $[1\overline{1}2]$ directions
surface normal and polished with 0.25~\micro\meter\ diamond paste to a
flat shiny surface.

The crystals were mounted in the ID08 5-circle diffractometer
operating at a base pressure of $1\times10^{-8}$~mbar, and
equipped with a Si diode detector. The beamline produces $\sim 100$ \% linearly polarized X-rays. At $\sim$650~eV an energy
resolution of 165~meV was obtained. Sample cooling was achieved using a liquid
He cryostat attached to the sample stage by copper braids
resulting in a base temperature of 63~K.

On the 5U1 beamline, crystals were mounted on a two circle
diffractometer enclosed in a high vacuum chamber with a base
pressure of $1\times10^{-8}$~mbar. The incident beam has a
resolution of 500~meV, and a beamsize of $1\times1$~mm$^{2}$. The sample was
mounted on a liquid nitrogen cooled copper block, with a small
manual $\chi$ adjustment for initial orientation.  The base
temperature achieved was 83~K. Temperature stability between 83~K
and 300~K was achieved with a heater element, and control
thermometer mounted next to the sample.

At both beamlines the experimental procedure was identical.  The
incident energy was set to the manganese $L_{\rm 3}$ edge, and superlattice
peaks located at the \oo\ and \mo\ positions.   Each reflection was
measured on a separate sample. Energy scans of the reflections
were performed at fixed wavevector.  The integrated intensity was
measured by longitudinal scans through the peak at fixed
energy.

The azimuthal dependence of the orbital order superlattice reflection was 
measured on the ID08 diffractometer by rotating the sample
around the scattering vector, $\vec{q}$. In the absence of an 
orientation matrix (UB) this was achieved by rotation of the diffractometer's 
$\phi$ axis. The  azimuthal angle $\Psi$ is  $90^{\circ}$  when
the $[001]$ direction lies within the scattering plane.

\subsection{ Theoretical model}

The calculations of the RXS spectra are based on  atomic multiplet 
calculations in a crystal field. Cowan's atomic multiplet 
program \cite{Cow68,Cow81,LaaTho91} provides {\it ab-initio} (Hartree-Fock) 
values of the radial Coulomb (Slater) integrals  $F^{0,2,4}(d,d)$, $F^2(p,d)$,
$G^{1,3}(p,d)$ (direct and exchange contributions) and the
spin-orbit interactions $\zeta(2p)$ and $\zeta(3d)$ for an isolated Mn$^{3+}$  ion.
Their values  are
given in table \ref{HartreeFockvalues}. As we are lowering
the symmetry by the inclusion of the crystal field, we apply the Wigner-Eckart
theorem to calculate matrix elements in a given point group \cite{But81} starting from the matrix elements
evaluated in the spherical group. This is implemented
in the ``Racah'' code, which also gives the values of the
dipole transition matrix elements, necessary for the calculation of the spectra, as will be
described in the following subsection.  The fitting procedure starts with adjusting the
crystal field type and strength by modifying the cubic (${\rm X^{400}}$) and
tetragonal (${\rm X^{220}}$) crystal field parameters. 
Orbital and magnetic spectra
calculated using the atomic multiplet code and including the
crystal field effects are then compared to the experimental ones and
the procedure is continued until the optimized set of crystal field
parameters is found.
In order to take into account the screening effects present in a real crystal
with respect to the atomic picture, we scale down all the Slater
integrals to 75~\% of their atomic values.

In the actual \lsmo\ compound, the spins are approximately aligned in the [110] direction
\footnote{There is experimental indication for the spin direction in the
[210]-direction \cite{sternlieb:2169}, but in our calculations we assumed
it was [110], as it simplified the scattering formula.},
displaying anti-ferromagnetic ordering along the $c$-axis 
and between the spin-chains, and ferromagnetic along the spin-chains, 
as shown in  Fig. \ref{fig:structure}c.
In order to simulate the effects of the spin ordering on an isolated ion,
(superexchange and direct exchange interactions)
we introduce a magnetic field acting on the spin of the atom. This field splits
additionally the $S=2$ quintet into $S_z=-2,-1, \dots,2$ levels. Inclusion of the magnetic 
field  favors the $S_z=-2$ level as the ground state of the atom. 
The strength of the exchange interaction  was set to 0.02~eV.

\renewcommand{\baselinestretch}{1}
\begin{table}[ht]
\bigskip
\begin{center}
\begin{tabular}{|l c c c c c c c| }
\hline
\hline
$ $ & $F^{2}(d,d)$  &  $F^{4}(d,d)$ &   $F^{2}(p,d)$ & $G^{1}(p,d)$ &$G^{3}(p,d)$  &$\zeta(2p)$ &$\zeta(3d)$    \\
\hline
$3d^4$  & $11.415$ & $7.148$ & $-$ & $-$ & $-$ & $-$ & $0.046$\\
$2p^53d^{5}$ &  $12.210$ & $7.649$  & $6.988$ & $5.179$ & $2.945$ & $7.467$ & $0.059$ \\
\hline
\hline
\end{tabular}
\caption{The Hartree-Fock values for the ground and excited states of $\rm Mn^{3+}$ given in eV. Coulomb and exchange integrals in the calculation have been scaled to 75~\% of their atomic values. The $p$-shell spin-orbit parameter $\zeta(2p)$ has been increased
by 9~\% from the Hartree-Fock value to correspond to the experimental value \cite{ThoAttGul01}. }
\label{HartreeFockvalues}
\end{center}
\end{table}

Using the above procedure one readily obtains
the eigenvalues for the ground ($3d^n$) and excited ($2p^53d^{n+1}$) states
separately. The next step  towards the calculation of the scattering or 
absorption spectrum is the evaluation of the dipole transition
probability. Depending on the crystal symmetry and the type of scattering
one is interested in, the angular dependence and the expectation value of the transition operators
of the resonant amplitude will vary \cite{CarTho94}.

The crystal field has the tetragonal ($D_{4h}$) symmetry. However, if  
the spin direction is assumed along [110],
the site symmetry is lowered to that of the $C_i$ point group.
Following the approach described by Carra and Thole \cite{CarTho94}, we derived the expression which 
describes the atomic resonant scattering amplitude \footnote{To be described in a future theoretical publication.}
and computed the structure factor.
In the case of the orbital scattering with the wave vector \oo,
the resonant scattering intensity is proportional to the following combination
of the atomic scattering tensor components:
\begin{equation}
\label{f_OO}
I^{\rm OO}_{\rm res} \propto| F^{\rm e}_{10;1-1}- 
F^{\rm e}_{10;11}+F^{\rm e}_{11;10}-F^{\rm e}_{1-1;10}|^2,
\end{equation}
with $ F^{\rm e}_{\rm 1m;1m'}$  defined as:
\begin{equation}
F^{\rm e}_{{\rm 1m;1m'}}=\sum_n \frac{\langle 0|J^{1\dag}_{\rm m}|n\rangle \langle n|J^{1}_{\rm m'}|0\rangle}{E_0-E_n+\hbar \omega + i\Gamma /2},
\end{equation}
where m and m' denote polarization states and $J^1_m$ are the electric dipole operators defined in spherical coordinates. $|0\rangle$ represents the ground state with energy $E_0$ and
$|n\rangle$ intermediate state with energy $E_n$. The photon energy is $\hbar \omega$  and $\Gamma$
stands for the broadening due to the core-hole lifetime. 

Similarly, for the magnetic scattering with the wave vector \mo, the scattering intensity can be expressed as:
\begin{equation}
\label{f_MO}
I_{\rm res}^{\rm MO} \propto 
|F^{\rm e}_{11;11}-F^{\rm e}_{1-1;1-1}|^2.
\end{equation}
This can be written using the more usual notation \cite{HanTraBlu88} in 
which $I_{\rm res}^{\rm MO} \propto |F^{\rm e}_{1,1}-F^{\rm e}_{1,-1}|^2$.

In these calculations we used $\Gamma = 0.5$~eV and in addition, the scattering intensity was convoluted with a Gaussian of width 0.1~eV to simulate the experimental (energy) resolution. 

\section{ Orbital and magnetic scattering}

\begin{figure}
  \begin{center}
  \includegraphics[width=0.9\columnwidth]{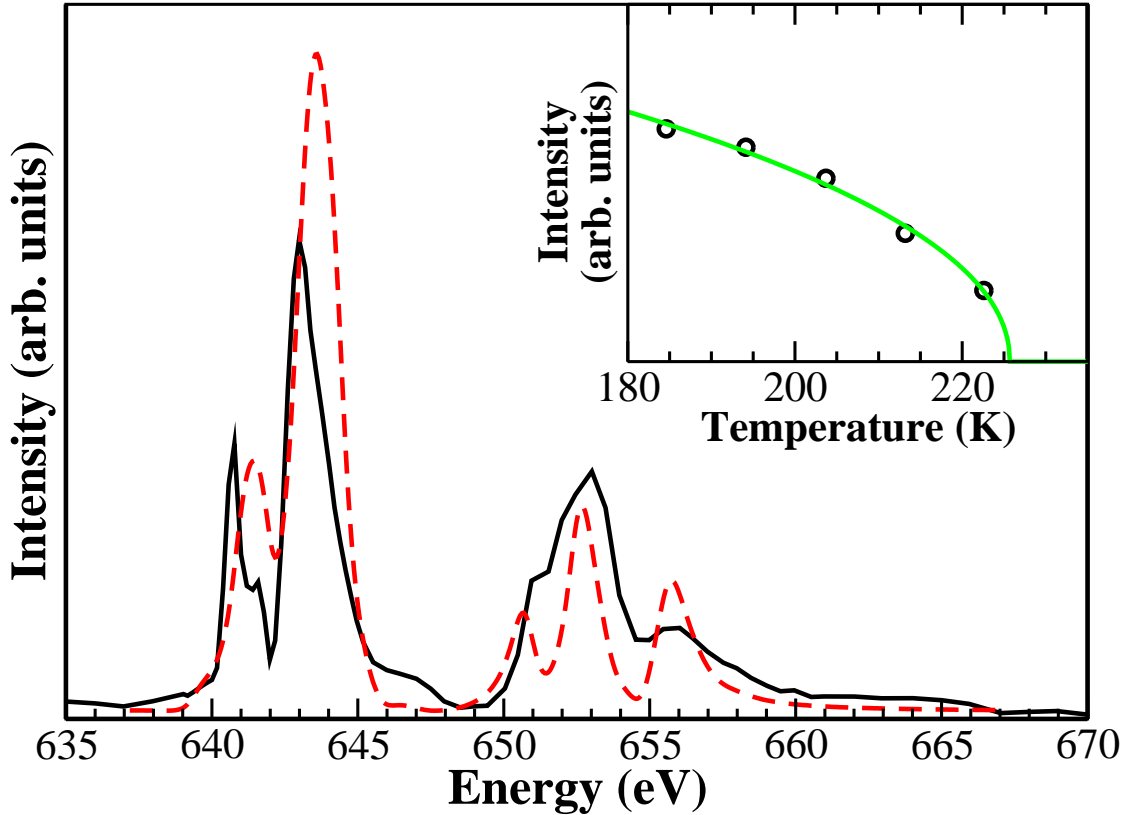}
  \caption{ Energy scan through the \oo\ orbital order reflection 
 at constant wavevector at 63~K (full black line) 
 with the theoretical fit (dashed red line).
 The inset shows the temperature evolution of the orbital order parameter. }
  \bigskip
  \label{fig:orb}
  \end{center}
\end{figure}
\begin{figure}
  \begin{center}
  \includegraphics[width=0.9\columnwidth]{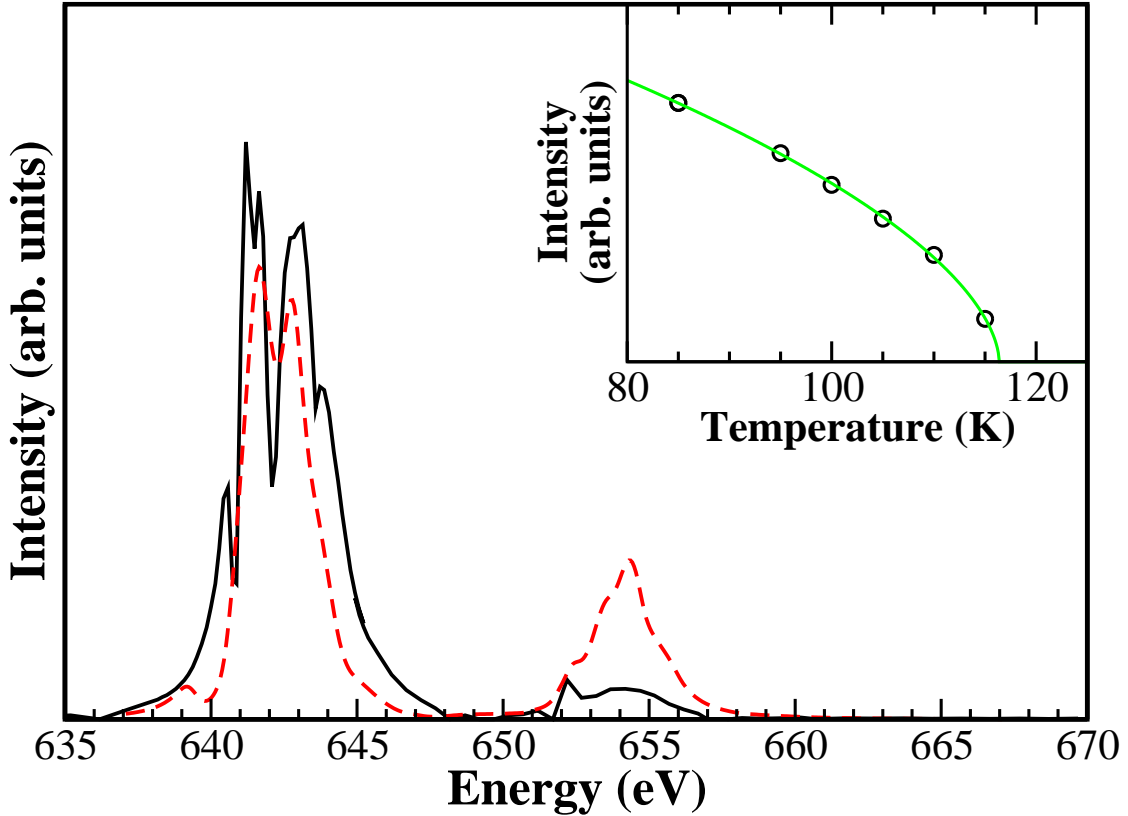}
  \caption{ Energy scan through the \mo\ magnetic order reflection 
at constant wavevector at 63~K (full black line) 
with the theoretical fit (dashed red line). The
inset shows the temperature evolution of the magnetic order parameter.}
  \bigskip
  \label{fig:magn}
  \end{center}
\end{figure}

Figure \ref{fig:orb}  shows the energy dependence of the scattered intensity at a fixed wavevector 
of $\vec{q}_{OO} =$\oo\ through the Mn $L_3$ and $L_2$ edges. 
This wavevector corresponds to the long range orbital ordering of the Mn$^{3+}$ ions 
within the $a-b$ plane. These data were collected at the beamline ID08 at the ESRF, 
Grenoble, France, on a single crystal cut with the [110] direction surface normal. 
On initial inspection, the spectrum is dominated by scattering at the $L_{3}$ edge where 
two distinct features are present. The scattering is weaker  at the $L_{2}$ edge where
three convoluted peaks are observed. Figure \ref{fig:magn} shows the energy dependence collected in a 
similar fashion, but at a wavevector of $\vec{q}_{AF} =$ \mo. This 
reflection corresponds to long range anti-ferromagnetic ordering on the Mn$^{3+}$ sub-lattice. 
The observation of a $c$-axis component in the wavevector of the anti-ferromagnetic reflection 
(compared to the orbital order wavevector) indicates that whilst the orbitals are ferro-ordered 
along the [001] direction, the magnetic moments are aligned anti-ferromagnetically. 
The magnetic ordering energy spectra is even more dominated by scattering at the $L_{3}$ edge, 
which is observed to contain up to five separate features. The $L_{2}$ features are much weaker 
and contain only two main peaks. These data show no relative shift in energy between magnetic and  
orbital spectra, at variance to what was reported by Thomas {\it et al.} \cite{ThoHilGre04}.
The insets of Figures \ref{fig:orb} and \ref{fig:magn}  
show the temperature evolution of the orbital and magnetic order parameters 
respectively. The orbital order parameter was found to decrease in a continuous fashion with increasing 
temperature in the vicinity of $T_{OO} = 230$~K. The anti-ferromagnetic reflection shows the same behavior, albeit, 
with the lower transition temperature of $T_{N} = 120$~K. 

\begin{figure}
  \begin{center}
  \includegraphics[width=0.9\columnwidth]{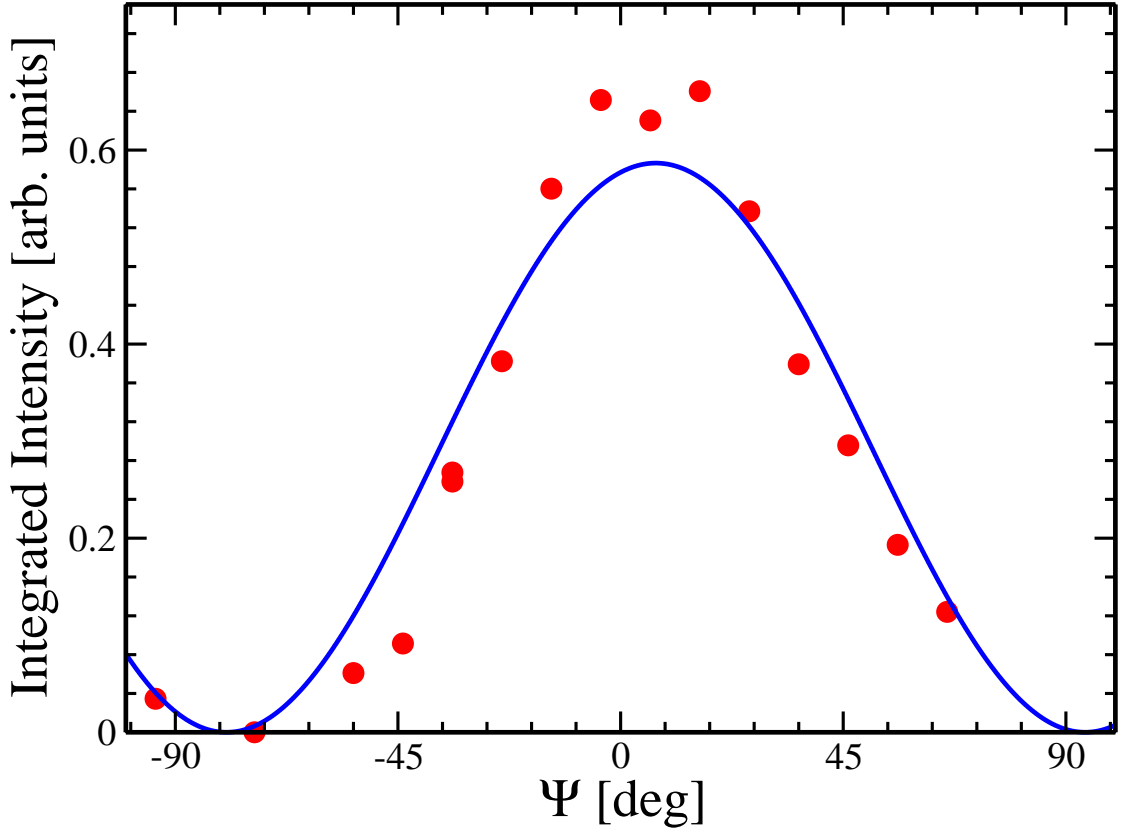}
  \caption{ Dependence of the integrated intensity as a function of azimuthal angle $\Psi$ 
  for the \oo\ superlattice reflection (circles). The solid line is a simulation of the 
  azimuthal dependence for the proposed orbital order.}
  \bigskip
  \label{fig:azimuth}
  \end{center}
\end{figure}

Figure~\ref{fig:azimuth} displays the integrated intensity of the orbital order superlattice reflection as a 
function of the azimuthal angle $\Psi$. The solid line shows a simulation of the azimuthal dependence 
given by the following equation
\begin{equation}
I(\theta,\psi) \propto \cos^{2}\theta_B \cos^{2} \Psi \cdot I_{\rm res}^{OO},
\end{equation}
where $\theta_B$ is the Bragg angle of the reflection and $\psi$ is the azimuthal angle. This dependence is 
characteristic of the $E1$ resonant scattering process. It is worth mentioning that this azimuthal dependence,
which shows an extinction of the scattering when $\Psi \approx 90^\circ$ is not consistent with scattering from
the magnetic structure with moments aligned along [210]. Therefore we conclude that the scattering observed at
the \oo\ position arises solely from the orbital ordering.

{In  Fig. \ref{fig:orb} we also present the best fit we have obtained for 
the orbital scattering by fitting simultaneously to the both data sets.}
The fitting procedure yielded the following optimized crystal field parameters: $ \rm X^{400}$ = 5.6~eV and 
$ \rm X^{220}$ = 3.75~eV, or 10$\rm D_q$ = 1.70~eV and $\rm D_s = -0.45$~eV . 
This corresponds to orbital ordering 
of the ${d_{x^2 - z^2}/ d_{y^2 - z^2}}$ type, as illustrated in Fig \ref{fig:structure}b. This is
in contrast to the conventional model of orbital ordering of the ${d_{3x^2 - r^2}/ d_{3y^2 - r^2}}$ type \cite{mizokawa:r493,castleton:1033,dhesi:056403}, but in agreement with the recent findings of Huang {\it et al.} \cite{HuaWuGuo04}.
Magnetic scattering corresponding to the
same crystal field parameters is shown in Fig. \ref{fig:magn}. All the features of the $L_{\rm 3}$ 
and $L_{\rm 2}$ structures of the orbital scattering fit display good agreement  
with the experimental data although the  $L_{\rm 3}/L_{\rm 2}$ branching ratio is 
somewhat overestimated, and the high-energy
shoulder of the $L_{\rm 3}$ peak is not reproduced.
 The magnetic scattering is also  well described, 
 but it overestimates the  intensity of the $L_{\rm 2}$ peak.  Both low- and high-energy shoulders in the
$L_{\rm 3}$ peak are not present in the fit. 
{We note that we could not obtain satisfactory simultaneous fits
to the orbital and magnetic scattering using the
other type of the orbital ordering (${d_{3x^2 - r^2}/d_{3y^2 - r^2}}$).
The fit corresponding to the crystal field described with $ \rm X^{400}$ = 5.6~eV and $ \rm X^{220}$ = -1.5~eV,
reproduced the spectrum for the orbital scattering very well (having all the peaks as the fit on Fig. \ref{fig:orb}
with even slightly improved  the $L_{\rm 3}/L_{\rm 2}$ ratio), but the magnetic scattering was poorly described,
completely missing the first feature in  the  $L_{\rm 3}$ peak.
We also note that the addition of a small orthorhombic ($D_{2h}$) component, $ \rm X^{222}$ = -0.85~eV, 
to the crystal field related to the ${d_{x^2 - z^2}/ d_{y^2 - z^2}}$ type of the orbital ordering
somewhat improved the $L_{\rm 3}/L_{\rm 2}$ ratio for the orbital scattering and 
broadened the previously underestimated width of the $L_{\rm 3}$ peak for the magnetic scattering, 
but kept the rest of the spectra shown in Figs. \ref{fig:orb} and \ref{fig:magn}  unchanged. }
 
\begin{figure}[ht]
  \begin{center}
  \includegraphics[width=0.8\columnwidth,angle=270]{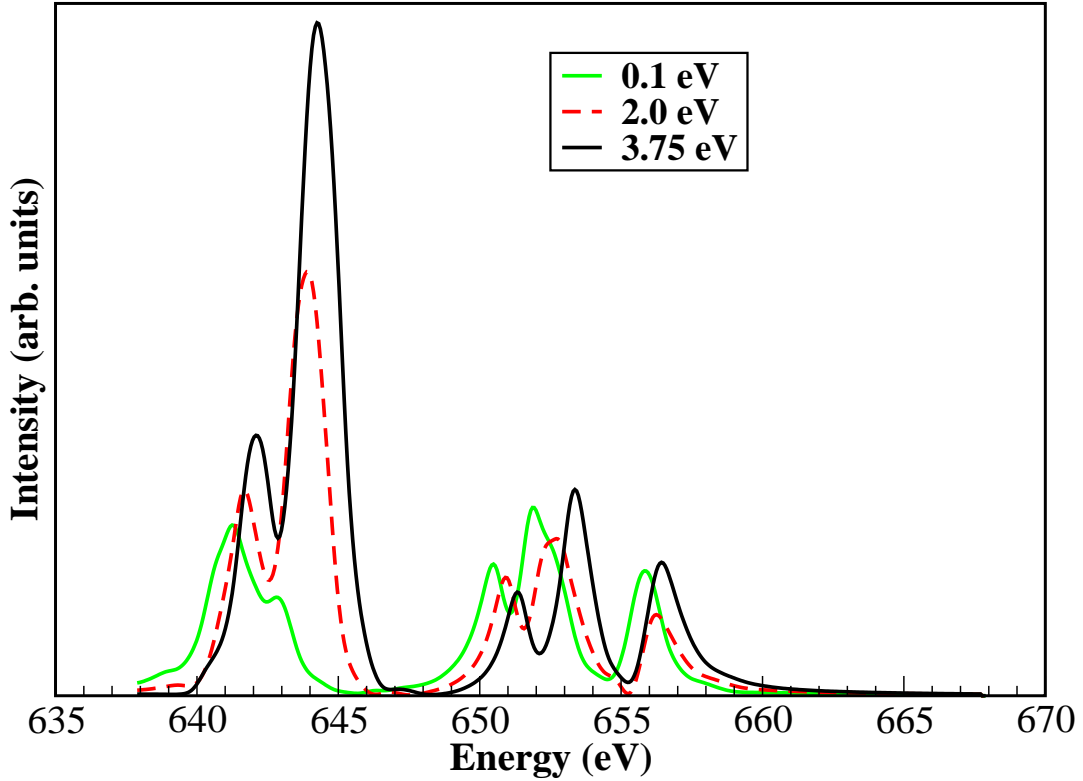}
  \caption{ Variation of the Jahn-Teller effect through the reduction of the tetragonal
crystal field: $\rm X^{220}$ = 3.75~eV (full black line), 2~eV (dashed red line) and 0.1~eV 
(full green line). }
  \bigskip
  \label{fig:JT}
  \end{center}
\end{figure}

In order to address the origin of the different spectral features, we have examined 
the effect of a reduction of the tetragonal part ($\rm X^{220}$) 
of the crystal field, using different values of $\rm X^{220}$  
as shown in Fig. \ref{fig:JT}. This approach is based on the fact that changes of the Jahn-Teller distortion
yield corresponding changes to the tetragonal field. 
We conclude that the reduction of the Jahn-Teller effect
results a reduced $ L_{\rm 3} / L_{\rm 2}$ ratio and in the drastic 
reduction of the main feature at the $L_{\rm 3}$ edge.

\section{Temperature dependence}

\begin{figure}
  \begin{center}
  \includegraphics[width=8.7cm]{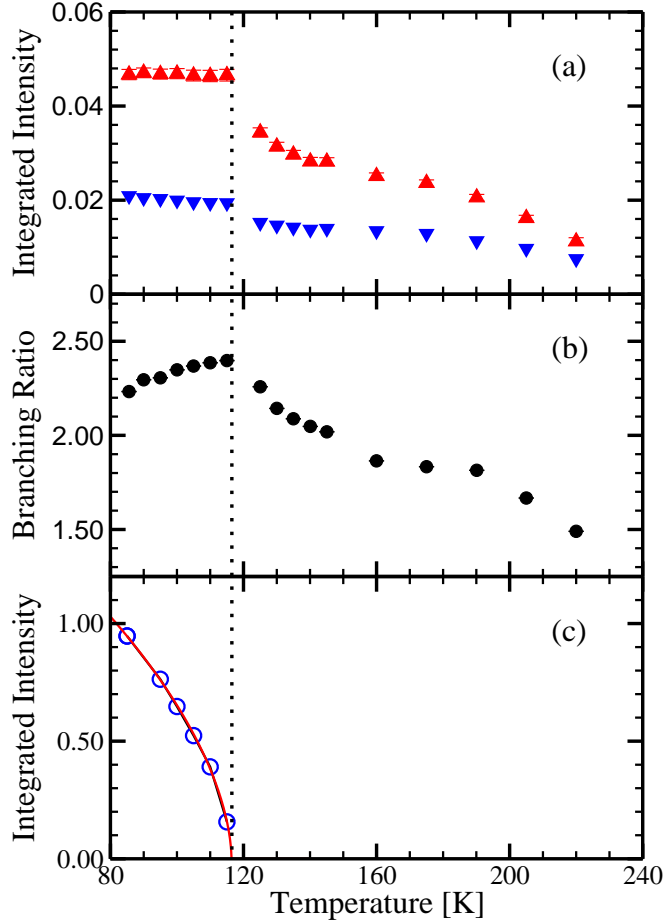}
  \caption{(a) Temperature dependence of the integrated intensity of the main
features at the $L_3$ edge (triangles) and at the $L_2$ edge (inverted triangles). 
(b) The temperature dependence of the ratio of the integrated intensities $L_3/L_2$.
(c) The temperature dependence of the integrated intensity of the \mo\ magnetic reflection 
measured at 643~eV.  }
  \bigskip
  \label{fig:temp}
  \end{center}
\end{figure}

 The orbital order spectrum shown in Fig.~\ref{fig:orb} was obtained at a
 temperature of 63~K. We also made measurements of spectral features in
 the \oo\ orbital order reflection as a function of temperature, upon
 warming, to measure the change of the crystal field.  These measurements
 were performed on station 5U1 and the energy resolution ($\Delta E$)
 was deliberately reduced to approximately integrate the spectral feature
 over energy. These results are displayed in Fig.~\ref{fig:temp}. 
 
Figure~\ref{fig:JT} shows that the scattered intensity at the $L_{3}$ edge is
very sensitive to the Jahn-Teller effect whilst the intensity at  the
$L_{2}$ edge is not. We define for the purposes of this discussion, the
branching ratio (Fig.~\ref{fig:temp} middle panel) as the ratio of integrated
intensities (over $\vec{q}$) measured at energies of 643~eV and 653~eV
for the $L_{3}$ and $L_{2}$ edges respectively, shown in Fig.~\ref{fig:temp} top panel. Therefore, this branching ratio gives an indication of the Jahn-Teller distortion. 
 
 Upon cooling past $T_{OO}$ the branching ratio increases,
suggesting an increase in the Jahn-Teller distortion. Approaching $T_{N}$
the integrated intensity for both spectral features rises sharply.
However, the branching ratio also increases at this point indicating
that the Jahn-Teller distortion becomes even more dominant. At $T_{N}$,
as indicated by the measured magnetic order parameter in the bottom panel
of Fig.~\ref{fig:temp}, both the integrated intensity and the branching ratio
saturate. Upon further cooling only a slight decrease in the branching
ratio is observed. 

At temperatures below $T_{N}$, the orbital order correlations are approximately constant even as the anti-ferromagnetic spin correlations are gradually increasing. The dramatic increase in the intensity of the orbital ordering superlattice reflection at $T_{N}$ is very unexpected. The stepwise increase in the integrated intensity suggests that the long-range orbital ordering interactions maximize at $T_{N}$. The temperature region between $T_{N}$ and $T_{OO}$ is one in which the orbital order correlations gradually increase with decreasing temperatures. These results demonstrate the strong interaction between the orbital and spin degrees of freedom. Is such behavior typical, or restricted to just \lsmo? Because our study is one of the first to directly measure both the orbital and magnetic correlations, and their temperature dependences, it is difficult to be certain. However, there are intriguing results obtained by other resonant X-ray studies at the $K$ edge of other $3d$ transition metal compounds that suggest that such effects may be a general feature of strongly correlated systems. In $\mathrm{LaMnO_{3}}$ RXS studies\cite{murakami:582} at the Mn $K$ edge observed scattering at the symmetry forbidden $(300)$ position. Such measurements, indirectly attributed to orbital ordering via Jahn-Teller distortions, showed that the intensity increases dramatically at $T_{OO}$ and thereafter remained almost constant until just above $T_{N}$ where the integrated intensity increased by 50\%. In $\mathrm{KCuF_{3}}$ a doubling of the integrated intensity of the orbital order reflection was observed upon cooling just above $T_{N}$\cite{paolasini:106403, caciuffo:174425}. What is surprising about all of these results is that the increase occurs well below the orbital ordering transition temperature where the orbital order parameter is expected to be saturated. A number of different causes of this effect has been proposed. Ishihara and Maekawa\cite{ishihara:R9252} suggested that such anomalous changes may signal a change in the type of orbital order. However, our energy resonance data do not show any dramatic changes in profile around $T_{N}$. Paolasini {\it et al.}\cite{paolasini:106403} suggested that the magnetic order is driven by the orbital order and that the spin exchange constants are determined by the relative orientation of the occupied orbitals. Alternatively, Binggeli and Altarelli\cite{binggeli:085117} have suggested that in $\mathrm{KCuF_{3}}$ the increase in the orbital order reflection at $T_{N}$ may simply suggest a low temperature structural transition. In \lsmo\ our results suggest that the intensity of the orbital order reflection at $T_{N}$ is linked to an increase in the Jahn-Teller distortion in the absence of any structural transitions.  
    
\section{Conclusions}
 
We have presented the energy dependence of the scattered intensity at a fixed wavevector 
of $\vec{q}_{OO} =$\oo\ and $\vec{q}_{AF} =$\mo\  through the Mn $L_{\rm 3}$ and $L_{\rm 2}$ edges. 

A good agreement was found between the experimental data and theoretical fits which included
a realistic description of the spin direction. A single
set of parameters describing the crystal field was used for both orbital and magnetic
reflections. It is noted that there was a slight  discrepancy in the branching ratio ($L_3/L_2$)
between data and theoretical results.  

{
The results of the fitting procedure demonstrate that the orbital ordering is predominantly of the
 $d_{x^2 - z^2}/ d_{y^2 - z^2}$ type. A satisfactory fit to the both data sets could not
be obtained with the other proposed type of the orbital ordering ($d_{3x^2-r^2}/ d_{3y^2-r^2}$).
The inclusion of a small orthorhombic component of the crystal field, 
in the case of the  $d_{x^2 - z^2}/ d_{y^2 - z^2}$ type of the
orbital ordering, moderately improved the fits.
Such results are consistent with the x-ray linear dichroism results of Huang {\it et al.}\cite{HuaWuGuo04}.
In addition, the x-ray scattering results confirm the existence of long-range order.}

Further theoretical investigation of the orbital scattering indicates that the intensity of
the $L_3$ edge is strongly sensitive to the Jahn-Teller distortion, whereas the $L_2$ edge
is mostly sensitive to orbital order. We believe this is a general feature which can be used as
a guideline in the interpretation of the orbital scattering in other manganite systems. 
We conclude therefore, in the specific case of \lsmo\ presented here, that the system is highly influenced by Jahn-Teller distortions. 

Measurements of the temperature dependence of the ratio of the integrated intensities
at the $L_3$ and $L_2$ edges show that there is a strong coupling between the
Jahn-Teller distortion and the long-range magnetic order. These results are not consistent
with a change of the orbital order type at $T_N$, as previously suggested \cite{IshMae98},
but are suggestive of an increase of the Jahn-Teller distortions below $T_N$.

\acknowledgements

This work was supported by the Synchrotron Radiation Related Theory Network, SRRTN, of the EU.
S. B. W. would like to thank the  European Commission for support in the frame of 
the `Training and Mobility of Researchers' program. N. S. gratefully acknowledges 
the assistance of Paolo Carra in learning how to use the Cowan and ``Racah'' codes.
P. D. H. would like to acknowledge support from the University of Durham Research 
Foundation. We are grateful for support from EPSRC for a studentship for T. A. W. B. 
and CLRC for access to the SRS and ESRF. 

\bibliography{mang}

\begin{thebibliography}{39}
\expandafter\ifx\csname natexlab\endcsname\relax\def\natexlab#1{#1}\fi
\expandafter\ifx\csname bibnamefont\endcsname\relax
  \def\bibnamefont#1{#1}\fi
\expandafter\ifx\csname bibfnamefont\endcsname\relax
  \def\bibfnamefont#1{#1}\fi
\expandafter\ifx\csname citenamefont\endcsname\relax
  \def\citenamefont#1{#1}\fi
\expandafter\ifx\csname url\endcsname\relax
  \def\url#1{\texttt{#1}}\fi
\expandafter\ifx\csname urlprefix\endcsname\relax\def\urlprefix{URL }\fi
\providecommand{\bibinfo}[2]{#2}
\providecommand{\eprint}[2][]{\url{#2}}

\bibitem[{\citenamefont{Osborne}(2000)}]{science}
\bibinfo{author}{\bibfnamefont{I.~S.} \bibnamefont{Osborne}},
  \bibinfo{journal}{Science} \textbf{\bibinfo{volume}{288}},
  \bibinfo{pages}{461} (\bibinfo{year}{2000}).

\bibitem[{\citenamefont{Radaelli et~al.}(1997)\citenamefont{Radaelli, Cox,
  Marezio, and Cheong}}]{radaelli:3015}
\bibinfo{author}{\bibfnamefont{P.~G.} \bibnamefont{Radaelli}},
  \bibinfo{author}{\bibfnamefont{D.~E.} \bibnamefont{Cox}},
  \bibinfo{author}{\bibfnamefont{M.}~\bibnamefont{Marezio}}, \bibnamefont{and}
  \bibinfo{author}{\bibfnamefont{S.-W.} \bibnamefont{Cheong}},
  \bibinfo{journal}{Phys. Rev. B} \textbf{\bibinfo{volume}{55}},
  \bibinfo{pages}{3015} (\bibinfo{year}{1997}).

\bibitem[{\citenamefont{Mutou and Kontani}(1999)}]{mutou:3685}
\bibinfo{author}{\bibfnamefont{T.}~\bibnamefont{Mutou}} \bibnamefont{and}
  \bibinfo{author}{\bibfnamefont{H.}~\bibnamefont{Kontani}},
  \bibinfo{journal}{Phys. Rev. Lett.} \textbf{\bibinfo{volume}{83}},
  \bibinfo{pages}{3685} (\bibinfo{year}{1999}).

\bibitem[{\citenamefont{Khomskii and van~den Brink}(2000)}]{khomskii:3329}
\bibinfo{author}{\bibfnamefont{D.}~\bibnamefont{Khomskii}} \bibnamefont{and}
  \bibinfo{author}{\bibfnamefont{J.}~\bibnamefont{van~den Brink}},
  \bibinfo{journal}{Phys. Rev. Lett.} \textbf{\bibinfo{volume}{85}},
  \bibinfo{pages}{3329} (\bibinfo{year}{2000}).

\bibitem[{\citenamefont{Hotta et~al.}(2001)\citenamefont{Hotta, Dagotto,
  Koizumi, and Takada}}]{hotta:2478}
\bibinfo{author}{\bibfnamefont{T.}~\bibnamefont{Hotta}},
  \bibinfo{author}{\bibfnamefont{E.}~\bibnamefont{Dagotto}},
  \bibinfo{author}{\bibfnamefont{H.}~\bibnamefont{Koizumi}}, \bibnamefont{and}
  \bibinfo{author}{\bibfnamefont{Y.}~\bibnamefont{Takada}},
  \bibinfo{journal}{Phys. Rev. Lett.} \textbf{\bibinfo{volume}{86}},
  \bibinfo{pages}{2478} (\bibinfo{year}{2001}).

\bibitem[{\citenamefont{Daoud-Aladine et~al.}(2002)\citenamefont{Daoud-Aladine,
  Rodriguez-Carvajal, Pinsard-Gaudart, Fernandez-Diaz, and
  Revcolevschi}}]{daoud-aladine:097205}
\bibinfo{author}{\bibfnamefont{A.}~\bibnamefont{Daoud-Aladine}},
  \bibinfo{author}{\bibfnamefont{J.}~\bibnamefont{Rodriguez-Carvajal}},
  \bibinfo{author}{\bibfnamefont{L.}~\bibnamefont{Pinsard-Gaudart}},
  \bibinfo{author}{\bibfnamefont{M.~T.} \bibnamefont{Fernandez-Diaz}},
  \bibnamefont{and}
  \bibinfo{author}{\bibfnamefont{A.}~\bibnamefont{Revcolevschi}},
  \bibinfo{journal}{Phys. Rev. Lett.} \textbf{\bibinfo{volume}{89}},
  \bibinfo{eid}{097205} (\bibinfo{year}{2002}).

\bibitem[{\citenamefont{Moritomo et~al.}(1995)\citenamefont{Moritomo, Tomioka,
  Asamitsu, Tokura, and Matsui}}]{moritomo:3297}
\bibinfo{author}{\bibfnamefont{Y.}~\bibnamefont{Moritomo}},
  \bibinfo{author}{\bibfnamefont{Y.}~\bibnamefont{Tomioka}},
  \bibinfo{author}{\bibfnamefont{A.}~\bibnamefont{Asamitsu}},
  \bibinfo{author}{\bibfnamefont{Y.}~\bibnamefont{Tokura}}, \bibnamefont{and}
  \bibinfo{author}{\bibfnamefont{Y.}~\bibnamefont{Matsui}},
  \bibinfo{journal}{Phys. Rev. B} \textbf{\bibinfo{volume}{51}},
  \bibinfo{pages}{3297} (\bibinfo{year}{1995}).

\bibitem[{\citenamefont{Sternlieb et~al.}(1996)\citenamefont{Sternlieb, Hill,
  Wildgruber, Luke, Nachumi, Moritomo, and Tokura}}]{sternlieb:2169}
\bibinfo{author}{\bibfnamefont{B.~J.} \bibnamefont{Sternlieb}},
  \bibinfo{author}{\bibfnamefont{J.~P.} \bibnamefont{Hill}},
  \bibinfo{author}{\bibfnamefont{U.~C.} \bibnamefont{Wildgruber}},
  \bibinfo{author}{\bibfnamefont{G.~M.} \bibnamefont{Luke}},
  \bibinfo{author}{\bibfnamefont{B.}~\bibnamefont{Nachumi}},
  \bibinfo{author}{\bibfnamefont{Y.}~\bibnamefont{Moritomo}}, \bibnamefont{and}
  \bibinfo{author}{\bibfnamefont{Y.}~\bibnamefont{Tokura}},
  \bibinfo{journal}{Phys. Rev. Lett.} \textbf{\bibinfo{volume}{76}},
  \bibinfo{pages}{2169} (\bibinfo{year}{1996}).

\bibitem[{\citenamefont{Goodenough}(1955)}]{goodenough:564}
\bibinfo{author}{\bibfnamefont{J.~B.} \bibnamefont{Goodenough}},
  \bibinfo{journal}{Phys. Rev.} \textbf{\bibinfo{volume}{100}},
  \bibinfo{pages}{564} (\bibinfo{year}{1955}).

\bibitem[{\citenamefont{Herrero-Martin
  et~al.}(2004)\citenamefont{Herrero-Martin, Garcia, Subias, Blasco, and
  S\'anchez}}]{HerGarSub04}
\bibinfo{author}{\bibfnamefont{J.}~\bibnamefont{Herrero-Martin}},
  \bibinfo{author}{\bibfnamefont{J.}~\bibnamefont{Garcia}},
  \bibinfo{author}{\bibfnamefont{G.}~\bibnamefont{Subias}},
  \bibinfo{author}{\bibfnamefont{J.}~\bibnamefont{Blasco}}, \bibnamefont{and}
  \bibinfo{author}{\bibfnamefont{M.~C.} \bibnamefont{S\'anchez}},
  \bibinfo{journal}{Phys. Rev. B, in the press; preprint at
  http://arxiv.org/abs/cond-mat/0406407}  (\bibinfo{year}{2004}).

\bibitem[{\citenamefont{Ferrari et~al.}(2003)\citenamefont{Ferrari, Towler, and
  Littlewood}}]{FerTowLit03}
\bibinfo{author}{\bibfnamefont{V.}~\bibnamefont{Ferrari}},
  \bibinfo{author}{\bibfnamefont{M.}~\bibnamefont{Towler}}, \bibnamefont{and}
  \bibinfo{author}{\bibfnamefont{P.~B.} \bibnamefont{Littlewood}},
  \bibinfo{journal}{Phys. Rev. Lett.} \textbf{\bibinfo{volume}{91}},
  \bibinfo{pages}{227202} (\bibinfo{year}{2003}).

\bibitem[{\citenamefont{Wohlan and Koehler}(1955)}]{wohlan:545}
\bibinfo{author}{\bibfnamefont{E.~O.} \bibnamefont{Wohlan}} \bibnamefont{and}
  \bibinfo{author}{\bibfnamefont{W.~C.} \bibnamefont{Koehler}},
  \bibinfo{journal}{Phys. Rev.} \textbf{\bibinfo{volume}{100}},
  \bibinfo{pages}{545} (\bibinfo{year}{1955}).

\bibitem[{\citenamefont{Murakami
  et~al.}(1998{\natexlab{a}})\citenamefont{Murakami, Kawada, Kawata, Tanaka,
  Arima, Moritomo, and Tokura}}]{murakami:1932}
\bibinfo{author}{\bibfnamefont{Y.}~\bibnamefont{Murakami}},
  \bibinfo{author}{\bibfnamefont{H.}~\bibnamefont{Kawada}},
  \bibinfo{author}{\bibfnamefont{H.}~\bibnamefont{Kawata}},
  \bibinfo{author}{\bibfnamefont{M.}~\bibnamefont{Tanaka}},
  \bibinfo{author}{\bibfnamefont{T.}~\bibnamefont{Arima}},
  \bibinfo{author}{\bibfnamefont{Y.}~\bibnamefont{Moritomo}}, \bibnamefont{and}
  \bibinfo{author}{\bibfnamefont{Y.}~\bibnamefont{Tokura}},
  \bibinfo{journal}{Phys. Rev. Lett.} \textbf{\bibinfo{volume}{80}},
  \bibinfo{pages}{1932} (\bibinfo{year}{1998}{\natexlab{a}}).

\bibitem[{\citenamefont{Mizokawa and Fujimori}(1997)}]{mizokawa:r493}
\bibinfo{author}{\bibfnamefont{T.}~\bibnamefont{Mizokawa}} \bibnamefont{and}
  \bibinfo{author}{\bibfnamefont{A.}~\bibnamefont{Fujimori}},
  \bibinfo{journal}{Phys. Rev. B} \textbf{\bibinfo{volume}{56}},
  \bibinfo{pages}{R493} (\bibinfo{year}{1997}).

\bibitem[{\citenamefont{Elfimov et~al.}(1999)\citenamefont{Elfimov, Anisimov,
  and Sawatzky}}]{elfimov:4264}
\bibinfo{author}{\bibfnamefont{I.~S.} \bibnamefont{Elfimov}},
  \bibinfo{author}{\bibfnamefont{V.~I.} \bibnamefont{Anisimov}},
  \bibnamefont{and} \bibinfo{author}{\bibfnamefont{G.~A.}
  \bibnamefont{Sawatzky}}, \bibinfo{journal}{Phys. Rev. Lett.}
  \textbf{\bibinfo{volume}{82}}, \bibinfo{pages}{4264} (\bibinfo{year}{1999}).

\bibitem[{\citenamefont{Benfatto et~al.}(1999)\citenamefont{Benfatto, Joly, and
  Natoli}}]{benfatto:636}
\bibinfo{author}{\bibfnamefont{M.}~\bibnamefont{Benfatto}},
  \bibinfo{author}{\bibfnamefont{Y.}~\bibnamefont{Joly}}, \bibnamefont{and}
  \bibinfo{author}{\bibfnamefont{C.~R.} \bibnamefont{Natoli}},
  \bibinfo{journal}{Phys. Rev. Lett.} \textbf{\bibinfo{volume}{83}},
  \bibinfo{pages}{636} (\bibinfo{year}{1999}).

\bibitem[{\citenamefont{Solovyev and Terakura}(1999)}]{solovyev:2825}
\bibinfo{author}{\bibfnamefont{I.~V.} \bibnamefont{Solovyev}} \bibnamefont{and}
  \bibinfo{author}{\bibfnamefont{K.}~\bibnamefont{Terakura}},
  \bibinfo{journal}{Phys. Rev. Lett.} \textbf{\bibinfo{volume}{83}},
  \bibinfo{pages}{2825} (\bibinfo{year}{1999}).

\bibitem[{\citenamefont{Mahadevan et~al.}(2001)\citenamefont{Mahadevan,
  Terakura, and Sarma}}]{mahadevan:066404}
\bibinfo{author}{\bibfnamefont{P.}~\bibnamefont{Mahadevan}},
  \bibinfo{author}{\bibfnamefont{K.}~\bibnamefont{Terakura}}, \bibnamefont{and}
  \bibinfo{author}{\bibfnamefont{D.~D.} \bibnamefont{Sarma}},
  \bibinfo{journal}{Phys. Rev. Lett.} \textbf{\bibinfo{volume}{87}},
  \bibinfo{eid}{066404} (\bibinfo{year}{2001}).

\bibitem[{\citenamefont{Benedetti et~al.}(2001)\citenamefont{Benedetti, van~den
  Brink, Pavarini, Vigliante, and Wochner}}]{benedetti:060408}
\bibinfo{author}{\bibfnamefont{P.}~\bibnamefont{Benedetti}},
  \bibinfo{author}{\bibfnamefont{J.}~\bibnamefont{van~den Brink}},
  \bibinfo{author}{\bibfnamefont{E.}~\bibnamefont{Pavarini}},
  \bibinfo{author}{\bibfnamefont{A.}~\bibnamefont{Vigliante}},
  \bibnamefont{and} \bibinfo{author}{\bibfnamefont{P.}~\bibnamefont{Wochner}},
  \bibinfo{journal}{Phys. Rev. B} \textbf{\bibinfo{volume}{63}},
  \bibinfo{eid}{060408} (\bibinfo{year}{2001}).

\bibitem[{\citenamefont{Ishihara and Maekawa}(1998)}]{IshMae98}
\bibinfo{author}{\bibfnamefont{S.}~\bibnamefont{Ishihara}} \bibnamefont{and}
  \bibinfo{author}{\bibfnamefont{S.}~\bibnamefont{Maekawa}},
  \bibinfo{journal}{Phys. Rev. Lett.} \textbf{\bibinfo{volume}{80}},
  \bibinfo{pages}{3799} (\bibinfo{year}{1998}).

\bibitem[{\citenamefont{Ohsumi et~al.}(2003)\citenamefont{Ohsumi, Murakami,
  Kiyama, Nakao, Kubota, and et~al.}}]{OhsMurKiy03}
\bibinfo{author}{\bibfnamefont{H.}~\bibnamefont{Ohsumi}},
  \bibinfo{author}{\bibfnamefont{Y.}~\bibnamefont{Murakami}},
  \bibinfo{author}{\bibfnamefont{T.}~\bibnamefont{Kiyama}},
  \bibinfo{author}{\bibfnamefont{H.}~\bibnamefont{Nakao}},
  \bibinfo{author}{\bibfnamefont{M.}~\bibnamefont{Kubota}}, \bibnamefont{and}
  \bibinfo{author}{\bibfnamefont{Y.~W.} \bibnamefont{et~al.}},
  \bibinfo{journal}{J. Phys. Soc. Jpn.} \textbf{\bibinfo{volume}{72}},
  \bibinfo{pages}{1006} (\bibinfo{year}{2003}).

\bibitem[{\citenamefont{Wilkins
  et~al.}(2003{\natexlab{a}})\citenamefont{Wilkins, Hatton, Roper, Prabhakaran,
  and Boothroyd}}]{wilkins:187201}
\bibinfo{author}{\bibfnamefont{S.~B.} \bibnamefont{Wilkins}},
  \bibinfo{author}{\bibfnamefont{P.~D.} \bibnamefont{Hatton}},
  \bibinfo{author}{\bibfnamefont{M.~D.} \bibnamefont{Roper}},
  \bibinfo{author}{\bibfnamefont{D.}~\bibnamefont{Prabhakaran}},
  \bibnamefont{and} \bibinfo{author}{\bibfnamefont{A.~T.}
  \bibnamefont{Boothroyd}}, \bibinfo{journal}{Phys. Rev. Lett.}
  \textbf{\bibinfo{volume}{90}}, \bibinfo{eid}{187201}
  (\bibinfo{year}{2003}{\natexlab{a}}).

\bibitem[{\citenamefont{Wilkins
  et~al.}(2003{\natexlab{b}})\citenamefont{Wilkins, Spencer, Hatton, Collins,
  Roper, Prabhakaran, and Boothroyd}}]{wilkins:167205}
\bibinfo{author}{\bibfnamefont{S.~B.} \bibnamefont{Wilkins}},
  \bibinfo{author}{\bibfnamefont{P.~D.} \bibnamefont{Spencer}},
  \bibinfo{author}{\bibfnamefont{P.~D.} \bibnamefont{Hatton}},
  \bibinfo{author}{\bibfnamefont{S.~P.} \bibnamefont{Collins}},
  \bibinfo{author}{\bibfnamefont{M.~D.} \bibnamefont{Roper}},
  \bibinfo{author}{\bibfnamefont{D.}~\bibnamefont{Prabhakaran}},
  \bibnamefont{and} \bibinfo{author}{\bibfnamefont{A.~T.}
  \bibnamefont{Boothroyd}}, \bibinfo{journal}{Phys. Rev. Lett.}
  \textbf{\bibinfo{volume}{91}}, \bibinfo{eid}{167205}
  (\bibinfo{year}{2003}{\natexlab{b}}).

\bibitem[{\citenamefont{Castleton and Altarelli}(2000)}]{castleton:1033}
\bibinfo{author}{\bibfnamefont{C.~W.~M.} \bibnamefont{Castleton}}
  \bibnamefont{and}
  \bibinfo{author}{\bibfnamefont{M.}~\bibnamefont{Altarelli}},
  \bibinfo{journal}{Phys. Rev. B} \textbf{\bibinfo{volume}{62}},
  \bibinfo{pages}{1033} (\bibinfo{year}{2000}).

\bibitem[{\citenamefont{Dhesi et~al.}(2004)\citenamefont{Dhesi, Mirone, Nadai,
  Ohresser, Bencok, Brookes, Reutler, Revcolevschi, Tagliaferri, Toulemonde
  et~al.}}]{dhesi:056403}
\bibinfo{author}{\bibfnamefont{S.~S.} \bibnamefont{Dhesi}},
  \bibinfo{author}{\bibfnamefont{A.}~\bibnamefont{Mirone}},
  \bibinfo{author}{\bibfnamefont{C.~D.} \bibnamefont{Nadai}},
  \bibinfo{author}{\bibfnamefont{P.}~\bibnamefont{Ohresser}},
  \bibinfo{author}{\bibfnamefont{P.}~\bibnamefont{Bencok}},
  \bibinfo{author}{\bibfnamefont{N.~B.} \bibnamefont{Brookes}},
  \bibinfo{author}{\bibfnamefont{P.}~\bibnamefont{Reutler}},
  \bibinfo{author}{\bibfnamefont{A.}~\bibnamefont{Revcolevschi}},
  \bibinfo{author}{\bibfnamefont{A.}~\bibnamefont{Tagliaferri}},
  \bibinfo{author}{\bibfnamefont{O.}~\bibnamefont{Toulemonde}},
  \bibnamefont{et~al.}, \bibinfo{journal}{Phys. Rev. Lett.}
  \textbf{\bibinfo{volume}{92}}, \bibinfo{eid}{056403} (\bibinfo{year}{2004}).

\bibitem[{\citenamefont{Thomas et~al.}(2004)\citenamefont{Thomas, Hill,
  Grenier, Kim, Abbamonte, Venema, Tomioka, Tokura, McMorrow, Sawatzky
  et~al.}}]{ThoHilGre04}
\bibinfo{author}{\bibfnamefont{K.~J.} \bibnamefont{Thomas}},
  \bibinfo{author}{\bibfnamefont{J.~P.} \bibnamefont{Hill}},
  \bibinfo{author}{\bibfnamefont{S.}~\bibnamefont{Grenier}},
  \bibinfo{author}{\bibfnamefont{Y.-J.} \bibnamefont{Kim}},
  \bibinfo{author}{\bibfnamefont{P.}~\bibnamefont{Abbamonte}},
  \bibinfo{author}{\bibfnamefont{L.}~\bibnamefont{Venema}},
  \bibinfo{author}{\bibfnamefont{A.~R.~Y.} \bibnamefont{Tomioka}},
  \bibinfo{author}{\bibfnamefont{Y.}~\bibnamefont{Tokura}},
  \bibinfo{author}{\bibfnamefont{D.~F.} \bibnamefont{McMorrow}},
  \bibinfo{author}{\bibfnamefont{G.}~\bibnamefont{Sawatzky}},
  \bibnamefont{et~al.}, \bibinfo{journal}{Phys. Rev. Lett.}
  \textbf{\bibinfo{volume}{92}}, \bibinfo{pages}{237204}
  (\bibinfo{year}{2004}).

\bibitem[{\citenamefont{Cowan}(1968)}]{Cow68}
\bibinfo{author}{\bibfnamefont{R.~D.} \bibnamefont{Cowan}},
  \bibinfo{journal}{J. Opt. Soc. Am.} \textbf{\bibinfo{volume}{58}},
  \bibinfo{pages}{808} (\bibinfo{year}{1968}).

\bibitem[{\citenamefont{Cowan}(1981)}]{Cow81}
\bibinfo{author}{\bibfnamefont{R.~D.} \bibnamefont{Cowan}},
  \emph{\bibinfo{title}{The Theory of Atomic Structure and Spectra}}
  (\bibinfo{publisher}{University of California Press, Berkeley},
  \bibinfo{year}{1981}).

\bibitem[{\citenamefont{van~der Laan and Thole}(1991)}]{LaaTho91}
\bibinfo{author}{\bibfnamefont{G.}~\bibnamefont{van~der Laan}}
  \bibnamefont{and} \bibinfo{author}{\bibfnamefont{B.~T.} \bibnamefont{Thole}},
  \bibinfo{journal}{Phys. Rev. B} \textbf{\bibinfo{volume}{43}},
  \bibinfo{pages}{13401} (\bibinfo{year}{1991}).

\bibitem[{\citenamefont{Butler}(1981)}]{But81}
\bibinfo{author}{\bibfnamefont{P.~H.} \bibnamefont{Butler}},
  \emph{\bibinfo{title}{Point Group Symmetry Applications}}
  (\bibinfo{publisher}{Plenum Press}, \bibinfo{address}{New York},
  \bibinfo{year}{1981}).

\bibitem[{\citenamefont{Thomson and et~al.}(2001)}]{ThoAttGul01}
\bibinfo{author}{\bibfnamefont{A.}~\bibnamefont{Thomson}} \bibnamefont{and}
  \bibinfo{author}{\bibnamefont{et~al.}}, \emph{\bibinfo{title}{X-Ray Data
  Booklet}} (\bibinfo{publisher}{Lawrence Berkeley National Laboratory,
  University of California}, \bibinfo{address}{Berkeley, CA 94720},
  \bibinfo{year}{2001}).

\bibitem[{\citenamefont{Carra and Thole}(1994)}]{CarTho94}
\bibinfo{author}{\bibfnamefont{P.}~\bibnamefont{Carra}} \bibnamefont{and}
  \bibinfo{author}{\bibfnamefont{B.~T.} \bibnamefont{Thole}},
  \bibinfo{journal}{Rev. of Mod. Phys.} \textbf{\bibinfo{volume}{66}},
  \bibinfo{pages}{1509} (\bibinfo{year}{1994}).

\bibitem[{\citenamefont{Hannon et~al.}(1988)\citenamefont{Hannon, Trammel,
  Blume, and Gibbs}}]{HanTraBlu88}
\bibinfo{author}{\bibfnamefont{J.~P.} \bibnamefont{Hannon}},
  \bibinfo{author}{\bibfnamefont{G.~T.} \bibnamefont{Trammel}},
  \bibinfo{author}{\bibfnamefont{M.}~\bibnamefont{Blume}}, \bibnamefont{and}
  \bibinfo{author}{\bibfnamefont{D.}~\bibnamefont{Gibbs}},
  \bibinfo{journal}{Phys. Rev. Lett.} \textbf{\bibinfo{volume}{61}},
  \bibinfo{pages}{1245} (\bibinfo{year}{1988}).

\bibitem[{\citenamefont{Huang et~al.}(2004)\citenamefont{Huang, Wu, Guo, Lin,
  Hou, Chang, Chen, Fujimori, Kimura, Huang et~al.}}]{HuaWuGuo04}
\bibinfo{author}{\bibfnamefont{D.~J.} \bibnamefont{Huang}},
  \bibinfo{author}{\bibfnamefont{W.~B.} \bibnamefont{Wu}},
  \bibinfo{author}{\bibfnamefont{G.~Y.} \bibnamefont{Guo}},
  \bibinfo{author}{\bibfnamefont{H.-J.} \bibnamefont{Lin}},
  \bibinfo{author}{\bibfnamefont{T.~Y.} \bibnamefont{Hou}},
  \bibinfo{author}{\bibfnamefont{C.~F.} \bibnamefont{Chang}},
  \bibinfo{author}{\bibfnamefont{C.~T.} \bibnamefont{Chen}},
  \bibinfo{author}{\bibfnamefont{A.}~\bibnamefont{Fujimori}},
  \bibinfo{author}{\bibfnamefont{T.}~\bibnamefont{Kimura}},
  \bibinfo{author}{\bibfnamefont{H.~B.} \bibnamefont{Huang}},
  \bibnamefont{et~al.}, \bibinfo{journal}{Phys. Rev. Lett.}
  \textbf{\bibinfo{volume}{92}}, \bibinfo{pages}{087202}
  (\bibinfo{year}{2004}).

\bibitem[{\citenamefont{Murakami
  et~al.}(1998{\natexlab{b}})\citenamefont{Murakami, Hill, Gibbs, Blume,
  Koyama, Tanaka, Kawata, Arima, Tokura, Hirota et~al.}}]{murakami:582}
\bibinfo{author}{\bibfnamefont{Y.}~\bibnamefont{Murakami}},
  \bibinfo{author}{\bibfnamefont{J.~P.} \bibnamefont{Hill}},
  \bibinfo{author}{\bibfnamefont{D.}~\bibnamefont{Gibbs}},
  \bibinfo{author}{\bibfnamefont{M.}~\bibnamefont{Blume}},
  \bibinfo{author}{\bibfnamefont{I.}~\bibnamefont{Koyama}},
  \bibinfo{author}{\bibfnamefont{M.}~\bibnamefont{Tanaka}},
  \bibinfo{author}{\bibfnamefont{H.}~\bibnamefont{Kawata}},
  \bibinfo{author}{\bibfnamefont{T.}~\bibnamefont{Arima}},
  \bibinfo{author}{\bibfnamefont{Y.}~\bibnamefont{Tokura}},
  \bibinfo{author}{\bibfnamefont{K.}~\bibnamefont{Hirota}},
  \bibnamefont{et~al.}, \bibinfo{journal}{Phys. Rev. Lett.}
  \textbf{\bibinfo{volume}{81}}, \bibinfo{pages}{582}
  (\bibinfo{year}{1998}{\natexlab{b}}).

\bibitem[{\citenamefont{Paolasini et~al.}(2002)\citenamefont{Paolasini,
  Caciuffo, Sollier, Ghigna, and Altarelli}}]{paolasini:106403}
\bibinfo{author}{\bibfnamefont{L.}~\bibnamefont{Paolasini}},
  \bibinfo{author}{\bibfnamefont{R.}~\bibnamefont{Caciuffo}},
  \bibinfo{author}{\bibfnamefont{A.}~\bibnamefont{Sollier}},
  \bibinfo{author}{\bibfnamefont{P.}~\bibnamefont{Ghigna}}, \bibnamefont{and}
  \bibinfo{author}{\bibfnamefont{M.}~\bibnamefont{Altarelli}},
  \bibinfo{journal}{Phys. Rev. Lett.} \textbf{\bibinfo{volume}{88}},
  \bibinfo{eid}{106403} (\bibinfo{year}{2002}).

\bibitem[{\citenamefont{Caciuffo et~al.}(2002)\citenamefont{Caciuffo,
  Paolasini, Sollier, Ghigna, Pavarini, van~den Brink, and
  Altarelli}}]{caciuffo:174425}
\bibinfo{author}{\bibfnamefont{R.}~\bibnamefont{Caciuffo}},
  \bibinfo{author}{\bibfnamefont{L.}~\bibnamefont{Paolasini}},
  \bibinfo{author}{\bibfnamefont{A.}~\bibnamefont{Sollier}},
  \bibinfo{author}{\bibfnamefont{P.}~\bibnamefont{Ghigna}},
  \bibinfo{author}{\bibfnamefont{E.}~\bibnamefont{Pavarini}},
  \bibinfo{author}{\bibfnamefont{J.}~\bibnamefont{van~den Brink}},
  \bibnamefont{and}
  \bibinfo{author}{\bibfnamefont{M.}~\bibnamefont{Altarelli}},
  \bibinfo{journal}{Phys. Rev. B.} \textbf{\bibinfo{volume}{65}},
  \bibinfo{pages}{174425} (\bibinfo{year}{2002}).

\bibitem[{\citenamefont{Ishihara and Maekawa}(2000)}]{ishihara:R9252}
\bibinfo{author}{\bibfnamefont{S.}~\bibnamefont{Ishihara}} \bibnamefont{and}
  \bibinfo{author}{\bibfnamefont{S.}~\bibnamefont{Maekawa}},
  \bibinfo{journal}{Phys. Rev. B} \textbf{\bibinfo{volume}{62}},
  \bibinfo{pages}{R9252} (\bibinfo{year}{2000}).

\bibitem[{\citenamefont{Binggeli and Altarelli}(2004)}]{binggeli:085117}
\bibinfo{author}{\bibfnamefont{N.}~\bibnamefont{Binggeli}} \bibnamefont{and}
  \bibinfo{author}{\bibfnamefont{M.}~\bibnamefont{Altarelli}},
  \bibinfo{journal}{Phys. Rev. B.} \textbf{\bibinfo{volume}{70}},
  \bibinfo{pages}{085117} (\bibinfo{year}{2004}).

\end{thebibliography}

\end{document}